\documentclass[11pt]{article}
\pdfoutput=1
\newcommand{\Comment}[1]{{}}
\usepackage{amsfonts,amsthm,amsmath,amssymb,slashed}
\usepackage[textwidth = 430 pt, textheight = 630 pt]{geometry}
\usepackage{color}
\usepackage{booktabs}

\Comment{\usepackage{color}
\definecolor{MyDarkBlue}{rgb}{0.15,0.15,0.45}
\usepackage[linktocpage=true]{hyperref}
\hypersetup{
colorlinks=true,
citecolor=MyDarkBlue,
linkcolor=MyDarkBlue,
urlcolor=MyDarkBlue,
pdfauthor={Heliudson Bernardo, Renato Costa, Horatiu Nastase and Amanda Weltman},
pdftitle={Conformal inflation with chameleon coupling},
pdfsubject={hep-th}
}
\usepackage[utf8]{inputenc}
\usepackage[numbers,sort&compress]{natbib}
\usepackage{hypernat}}
\usepackage{graphicx}
\usepackage{cite}
\usepackage[colorlinks=true]{hyperref}
\hypersetup{
  allcolors = blue,
}

\newcommand\ignore[1]{}
\def\one{{\,\hbox{1\kern-.8mm l}}}

\def\a{\alpha}

\def\d{\partial}

\newcommand{\Cset}{{\,\,{{{^{_{\pmb{\mid}}}}\kern-.45em{\mathrm C}}}}}

\newcommand{\be}{\begin{equation}}
\newcommand{\bea}{\begin{eqnarray}}

\newcommand{\ee}{\end{equation}}
\newcommand{\eea}{\end{eqnarray}}

\begin{document}

\renewcommand{\thefootnote}{\fnsymbol{footnote}}

\makeatletter
\@addtoreset{equation}{section}
\makeatother
\renewcommand{\theequation}{\thesection.\arabic{equation}}

\rightline{}
\rightline{}




\begin{center}
{\LARGE \bf{\sc Conformal inflation with chameleon coupling}}
\end{center}
 \vspace{1truecm}
\thispagestyle{empty} \centerline{
{\large \bf {\sc Heliudson Bernardo${}^{a},$}}\footnote{E-mail address: \Comment{\href{mailto:heliudson@ift.unesp.br}}{\tt heliudson@ift.unesp.br}}
{\large \bf {\sc Renato Costa${}^{b},$}}\footnote{E-mail address: \Comment{\href{mailto:Renato.Santos@uct.ac.za}}{\tt Renato.Santos@uct.ac.za}}
{\large \bf {\sc Horatiu Nastase${}^{a}$}}\footnote{E-mail address: \Comment{\href{mailto:nastase@ift.unesp.br}}{\tt nastase@ift.unesp.br}}
{\bf{\sc and}}
{\large \bf {\sc Amanda Weltman${}^{b}$}}\footnote{E-mail address: \Comment{\href{mailto:amanda.weltman@uct.ac.za}}{\tt amanda.weltman@uct.ac.za}}
                                                        }

\vspace{.5cm}


\centerline{{\it ${}^a$Instituto de F\'{i}sica Te\'{o}rica, UNESP-Universidade Estadual Paulista}}
\centerline{{\it R. Dr. Bento T. Ferraz 271, Bl. II, Sao Paulo 01140-070, SP, Brazil}}
\vspace{.3cm}
\centerline{{\it ${}^b$The Cosmology and Gravity Group, Department of Mathematics and Applied Mathematics,}}
\centerline{{\it University of Cape Town, Private Bag, Rondebosch, 7700, South Africa}}

\vspace{1truecm}

\thispagestyle{empty}

\centerline{\sc Abstract}

\vspace{.4truecm}

\begin{center}
\begin{minipage}[c]{380pt}
{\noindent We investigate the possibility that the inflaton, in particular in conformal inflation models, is also a chameleon, i.e. that it couples to
the energy density of some heavy non-relativistic matter present during inflation. We find new and interesting attractor-like behaviours, either prolonging
inflation, or changing the observables $n_s,r$, depending on the sign of the chameleon coupling exponent. We also check that the chameleon coupling with the heavy matter field strongly suppress entropy modes during inflation.
}
\end{minipage}
\end{center}

\vspace{.5cm}

\setcounter{page}{0}
\setcounter{tocdepth}{2}

\newpage

\renewcommand{\thefootnote}{\arabic{footnote}}
\setcounter{footnote}{0}

\linespread{1.1}
\parskip 4pt



\section{Introduction}

The idea that a scalar can have a ``chameleon" coupling to the (non-relativistic) matter density was introduced, in part, to allow for 
a scalar that can be very light on cosmological scales while also "hiding" its effects from observations in the lab (on Earth), 
or in the Solar System \cite{Khoury:2003rn,Khoury:2003aq}. Various laboratory searches have been initiated for such a scalar
(e.g.,\cite{Brax:2004qh,Upadhye:2006vi,Burrage:2007ew,Brax:2007vm,Brax:2007hi,Chou:2008gr,Weltman:2008fp,Burrage:2008ii,Upadhye:2009iv, Brax:2009aw, Steffen:2009sc,Brax:2009bk, Brax:2009ey, Steffen:2010ze,Brax:2013cfa,Armengaud:2014gea,Hamilton:2015zga,Elder:2016yxm}).
From a theoretical perspective, this alleviates the problem of having too many a priori light scalars in string theory (moduli): 
if they are chameleons, they don't contradict known experiments to date.
A way to embed chameleons in string theory was suggested in \cite{Hinterbichler:2010wu}.

On the other hand, since the chameleon is a scalar, an economical ansatz is for the same field that acts as an inflaton near the 
Big Bang to be the chameleon. This idea was explored in \cite{Hinterbichler:2013we}. In this case, however, the two regions (inflation and
chameleon) are separated by a large region of vanishing potential in field space, and the inflationary era itself is not affected 
by the chameleon coupling. An attempt to consider what happens if we consider inflation in the presence of a chameleon or symmetron
\cite{Hinterbichler:2010es} coupling was considered in \cite{Dong:2013swa} and \cite{Brax:2014baa}.

In this paper we want to consider the issue of inflation with a chameleon coupling taking into account that there can be new  ``attractor-like" phases
due to the chameleon coupling, where various forms of matter (contributions to the energy-momentum tensor) scale in the same way with
the scale factor, as seen for instance in  \cite{Hinterbichler:2013we} at zero potential. Since we need to consider  ``new inflation" type of models
with a plateau, a natural starting point is the system of  ``conformal inflation" models (see for instance \cite{Kallosh:2013hoa,Kallosh:2013maa}),
as analyzed in \cite{Costa:2014lta}.

We will assume the existence of some heavy, non-relativistic matter with density $\rho_X$ during the plateau phase (inflation), 
coupled to the inflaton via a chameleon coupling, $\rho_X F(\phi)$, and for the coupling the standard form $F=e^{c\phi/M_{\rm Pl}}$. We will investigate the
possibility of attractor-like behaviour due to this coupling, and see that depending on the sign of $c$, 
we can have either a prolonged period of attractor behaviour before inflation, 
or an effective inflationary potential that is different, with modified values for the CMBR observables $n_s,r$.

The paper is organized as follows. In section 2 we start by describing the set-up of conformal inflation coupled to the energy density 
through a chameleon coupling, and then derive the equations of motion, and note that depending on the sign of $c$ appearing in $F(\phi)$, we have
two different cases. In section 3 we describe the evolution of the system, and note that it leads to two attractors in the $c>0$ case and a
single one in the $c<0$ case. In section 4 we consider the modifications the chameleon coupling implies for the inflationary region, and we find
that in the $c<0$ it shortens inflation, whereas in the $c>0$ case it modifies it, leading to different values for the CMBR observables $n_s$ and $r$.
In section 5 we conclude.


\section{Conformal inflation and set-up}

\subsection{Conformal inflation coupled to energy density}

Conformal inflation is a perturbation of an exactly ``conformally invariant" model, that doesn't contain a fundamental scale, even the Planck
scale; all the scales appear from gauge fixing and minimizing the potential\footnote{See \cite{deCesare:2016mml} for a 
generalization of this approach in a Weyl geometry set up.}. The action is
\be
S=\frac{1}{2}\int d^4x\sqrt{-g}\left[\d_\mu\chi\d^\mu\chi-\d_\mu\phi\d^\mu\phi+\frac{\chi^2-\phi^2}{6}R-\frac{\lambda}{18}(\phi^2-\chi^2)^2\right].
\ee
Notice that the coupling of the scalars to the Einstein term has the conformal value, and that the potential is quartic, such as not to have any
dimensionful parameters in the action.

Then we have a local ``Weyl" type symmetry, acting on the fields by
\be
g_{\mu\nu}\rightarrow e^{-2\sigma(x)}g_{\mu\nu};\;\;\;
\chi\rightarrow e^{\sigma(x)}\chi, \;\;\;\;\;
\phi\rightarrow e^{\sigma(x)}\phi.
\ee
We also have an $SO(1,1)$ symmetry rotating $(\phi,\chi)$ (note that we have only the combination $\chi^2-\phi^2$ and the corresponding kinetic
term), which acts as a Lorentz type symmetry (originally, in \cite{Bars:2011mh,Bars:2011aa,Bars:2012mt}, the
model was motivated by 2-time physics: a covariant $(4,2)$-dimensional form led to it, with the $SO(1,1)$ being
a remnant of the $SO(4,2)$ Lorentz invariance).
Alternatively, the $SO(1,1)$ symmetry can also be obtained from a model with $SO(4,2)$ conformal invariance in 3+1 dimensions, as motivated in
 \cite{Ferrara:2010in,Kallosh:2013lkr,Kallosh:2013pby}.

 The field $\chi$ has the wrong sign kinetic term, so it would seem it is a ghost. However, the local ``Weyl" symmetry above allows 
 one to use a gauge choice to set it to zero, therefore the ghost is not physical. Choosing a gauge for the local ``Weyl" symmetry also introduces a scale,
the Planck scale, that will appear in front of the Einstein action. Yet, since the theory has only a single scale, its value is simply a definition of units, 
defining for instance what ``one meter" is (or what is $10^{19}GeV$), and physics is independent of the value we attribute to this scale,
since it is a gauge choice.

The gauge we will be mostly interested in is the Einstein gauge, defined by $\chi^2-\phi^2=6M_{\rm Pl}^2$. We solve this constraint in terms of a
{\em canonically normalized field} $\varphi$ by
\be
\chi=\sqrt{6}M_{\rm Pl}\cosh\frac{\varphi}{\sqrt{6}M_{\rm Pl}};\;\;\;
\phi=\sqrt{6}M_{\rm Pl}\sinh\frac{\varphi}{\sqrt{6}M_{\rm Pl}}.\label{chiphivarphi}
\ee
In terms of $\varphi$, we obtain the simple Einstein plus canonical scalar action, with a cosmological constant,
\be
S=\int d^4x\sqrt{-g}\left[\frac{M_{\rm Pl}^2}{2}R-\frac{1}{2}\d_\mu\varphi\d^\mu\varphi-\lambda M_{\rm Pl}^4\right].\label{einstein}
\ee

To obtain an inflationary model from the one above, we must deform the theory so that the cosmological constant gets modified into a potential
with a plateau. 
We do so by keeping the local ``Weyl" symmetry, and deforming the $SO(1,1)$ symmetry so that it is only approximately valid,
at large field values. Imposing the ``Weyl" invariance, we must have a potential of the type $f(\phi/\chi)\phi^4$, since both $\phi/\chi$ and
$\sqrt{-g}\phi^4$ are locally ``Weyl" invariant. Imposing moreover that the potential reduces to the $(\phi^2-\chi^2)^2$ form at large field values,
we finally obtain the most general form
\be
V=\lambda\left[\tilde f(\phi/\chi)\phi^2-\tilde g(\phi/\chi)\chi^2\right]^2\;,\label{fgtilde}
\ee
or in another parametrization
\be
V=\lambda f(\phi/\chi)\left[\phi^2-g(\phi/\chi)\chi^2\right]^2.
\ee
In the last form, in order to have the $SO(1,1)$ symmetry at large field values, since
in the Einstein gauge $\phi/\chi=\tanh \varphi/\sqrt{6}M_{\rm Pl}$, which goes to 1 at large $\varphi$, we must impose $g(1)=1$.

For simplicity, we will consider only cases with $f(x)=1$. Moreover, as in \cite{Costa:2014lta}, with a simple polynomial form for $g(x)$,
\be
g(x)=\omega^2+(1-\omega^2)x^n\;,\label{gofx}
\ee
where $\omega=246 GeV/\sqrt{6}M_{\rm Pl}$ and $n>2$,
we can interpolate between the conformal inflation plateau and a Higgs potential at small field values,
$V\simeq \left[\varphi^2-6\omega^2M_{\rm Pl}^2\right]^2$, so it presents a simple set-up, with the same scalar playing the role of inflaton and
Higgs. In this case, at $\varphi\rightarrow\infty$, using that $\omega^2\ll 1$, we obtain the potential
\be
V\simeq 9(n-2)^2\lambda M_{\rm Pl}^4\left[1-2n e^{-\sqrt{\frac{2}{3}}\frac{\varphi}{M_{\rm Pl}}}\right].\label{vinfty}
\ee
This exponentially-corrected plateau behaviour is the generic one for any function $g(x)$ that is well behaved near $x=1$, where we have the
large-field (large $\varphi$) expansion.

One easily find the scalar spectral index and the tensor to scalar ratio, $n_s$ and $r$, in terms of the number of e-folds $N_e$. In this case
\bea
1-n_s&\simeq&\frac{2}{N_e}\cr
r&\simeq&3(n_s-1)^2\simeq\frac{12}{N_e^2}\;,
\eea
as in the Starobinsky model. Thus the simplest model of conformal inflation effectively gives the Starobinsky model.
More generally, the Starobinsky model result above is found from the general potential at $\varphi\rightarrow\infty$
\be
V\simeq A\left[1-Be^{-\sqrt{\frac{2}{3}}\frac{\varphi}{M_{\rm Pl}}}\right].\label{generalpot}
\ee
If one replaces $\sqrt{2/3}$ in the exponent by a general factor $a$, $n_s$ is unchanged, but $r$ is changed to $8/(a^2N_e^2)$.

In the class of {\em conformal inflation} models, as we saw, any generic, well behaved at $x=1$, function $g(x)$ will give the simple
Starobinsky model at large field $\varphi$. Yet, with a slightly more unusual function $g(x)$, which can be defined implicitly by
\bea
V(\varphi)&=&36\lambda M_{\rm Pl}^4\sinh^4\frac{\varphi}{\sqrt{6}M_{\rm Pl}}
\left[1-\frac{g\left(\tanh\frac{\varphi}{\sqrt{6}M_{\rm Pl}}\right)}{\tanh^2\frac{\varphi}{\sqrt{6}M_{\rm Pl}}}\right]^2\cr
&\rightarrow&\frac{9}{4} \lambda M^4_{\rm Pl}\left(1-4e^{-\sqrt{\frac{2}{3}}\frac{\varphi}{M_{\rm Pl}}}\right)e^{2\sqrt{\frac{2}{3}}\frac{\varphi}{M_{\rm Pl}}}
\left[1-\frac{g\left(1-2e^{-\sqrt{\frac{2}{3}}\frac{\varphi}{M_{\rm Pl}}}\right)}{1-4e^{-\sqrt{\frac{2}{3}}\frac{\varphi}
{M_{\rm Pl}}}}\right]^2\;,
\eea
we can also obtain the potential
\be
V(\varphi)\simeq A\left[1-B\left(\frac{\varphi}{M_{\rm Pl}}\right)^{-p}\right]\;,\label{invpower}
\ee
at least approximately at large $\varphi$. Specifically, we need
\be
g\left(1-2e^{-\sqrt{\frac{2}{3}}x}\right)\simeq 1-4e^{-\sqrt{\frac{2}{3}}x}+e^{-2\sqrt{\frac{2}{3}}x}\left(1-\frac{B}{2}x^{-p}\right)\;,\;\;\;\; x\gg 1\;,
\ee
or equivalently, near $y\simeq 1$,
\be
g(y)\simeq y^2+\left(\frac{1-y}{2}\right)^2\left(1- \frac{B}{2}\left[-\sqrt{\frac{3}{2}}\ln \frac{1-y}{2}\right]^{-p}\right).
\ee
The above potential is interesting because it combines the necessary inflationary plateau (the constant
term) with the inverse power law potential which was the first model for a chameleon.

Consider that the canonical scalar $\varphi$ is not just the inflaton as above, but is also a chameleon, coupling universally to the
(non-relativistic) energy density $\rho$ through a coupling $\rho F(\varphi)$, where
\be
F(\varphi)=e^{-\frac{c}{M_{\rm Pl}}(\varphi-\varphi_0)}\;,
\ee
and where $\varphi_0\gg M_{\rm Pl}$ is some large VEV introduced to normalize the energy density when $\varphi=\varphi_0$.

The field $\varphi$ being a chameleon, means that the above coupling is relevant whenever there is a non-relativistic energy density.
That is certainly true during the current matter (and $\Lambda$) dominated phase, the period for which the chameleon was invented.

However, it will also be relevant in the case that there is a very heavy non-relativistic species $X$ during inflation and before, with a decay time scale
$\tau=1/\Gamma$ of the order of (or larger) than the time scale for inflation. Such a species will couple to the chameleon through its
energy density $\rho_m$, for a total energy density $\rho_X=\rho_m F(\varphi)$. While the existence of such a species $X$ might seem arbitrary,
certainly at the Planck scale, we expect that there will be many massive species, so it is not a stretch to assume such a species will exist also during inflation.
As for it being non-relativistic, we should remember that in slow-roll inflation also one assumes kinetic energy much smaller than the potential one.
Extending this assumption to $X$ would mean kinetic energy much smaller than the mass term (rest energy), leading to a non-relativistic species.
Then moreover the natural scale for $\Gamma$ is also of the order of $H$ during inflation. In conclusion, our assumption is certainly special, but
not more so than inflation itself.

\subsection{Equations of motion and two cases}

The equations of motion for the cosmology are the Friedmann equation for gravity, together with the energy conservation equation or
equivalently the acceleration equation, and the Klein-Gordon (KG) equation for the scalar. We will write them as an evolution in 
terms of the number of e-folds, $N$, defined by $dN=d\ln a$, instead of as a time evolution. Doing so simplifies the analysis.

In terms of $N$, the non-relativistic, chameleon-coupled energy density scales as
\be
\rho_X=F(\varphi)\rho_m=\frac{F(\varphi)\rho_0 a_0^3}{a^3}\propto e^{-3N}F(\varphi).
\ee
and radiation scales as
\be
\rho_{\rm rad}=\frac{\rho_{\rm rad,0}a_0^4}{a^4}\propto e^{-4N}.
\ee

The KG equation for the scalar is now
\be
\ddot \varphi+3H\dot \varphi=-\frac{\rho_0 a_0^3}{a^3}\frac{dF}{d\varphi} - \frac{dV}{d\varphi}=-\frac{\rho_X}{F    }\frac{dF}{d\varphi}- \frac{dV}{d\varphi}.
\ee

To write the Friedmann and acceleration equations for gravity, note first that the
energy-momentum tensor of a homogeneous scalar with canonical kinetic term and potential $V$ is
\be
T_{\mu\nu}=\d_\mu\varphi\d_\nu\varphi-\frac{1}{2}g_{\mu\nu}[(\d\varphi)^2+2V]\;,
\ee
which means that energy density and pressure associated with it are
\be
\rho_\varphi=\frac{\dot\varphi^2}{2}+V;\;\;\; p_\varphi=\frac{\dot\varphi^2}{2}-V.
\ee

The Friedmann equation is
\be
3M_{\rm Pl}^2H^2=\rho_{\rm tot}=\rho_X +\frac{\dot\varphi^2}{2}+\rho_{\rm rad}+V.
\ee

To simplify our notation going forward, we will denote $\frac{d}{dN}\equiv\;\; '$. Using
\be
\varphi'=\frac{d\varphi}{dt}\frac{dt}{d\ln a}=\frac{\dot \varphi}{H}\;,
\ee
we rewrite the Friedmann equation as
\be
3M_{\rm Pl}^2H^2=\rho_X+H^2\varphi'^2+\rho_{\rm rad}+V\;,
\ee
which gives
\be
3M_{\rm Pl}^2H^2=\frac{\rho_X+\rho_{\rm rad}+V}{1-\frac{\varphi'^2}{6M_{\rm Pl}^2}}.\label{friedmann}
\ee

The acceleration equation is
\be
\frac{\ddot a}{a}=H\frac{dH}{dN}+H^2=-\frac{(\rho_{\rm tot}+3p_{\rm tot})}{6M_{\rm Pl}^2}
=-\frac{1}{6M_{\rm Pl}^2}(\rho_X+2\rho_{\rm rad}+2\dot\varphi^2-2V)\;,
\ee
where as usual $\cdot= \frac{d}{dt}$. 
Now we can rewrite the KG equation first as
\be
H^2\varphi''+(3H^2+HH')\varphi'=-\frac{\rho_X}{F}\frac{dF(\varphi)}{d\varphi}-\frac{dV}{d\varphi}\;,
\ee
and using the acceleration and the Friedmann equation to calculate the friction term (the bracket in front of $\varphi'$), we find
the $N$-dependent form
\be
H^2\varphi''+\frac{1}{3M_{\rm Pl}^2}\left(\frac{3}{2}\rho_X+\rho_{\rm rad}+3V\right)\varphi'=-\frac{\rho_X}{F}\frac{dF}{d\varphi}
-\frac{dV}{d\varphi}=-\frac{dV_{\rm eff}}{d\varphi}\;,\label{KGV}
\ee
where
\be
V_{\rm eff}=V+\rho_X=V+F(\varphi)\rho_{m,0}e^{-3N}\;,
\ee
and the second term scales as $e^{-\frac{c(\varphi-\varphi_0)}{M_{\rm Pl}}-3N}$.

We can define the usual ratios of the densities to the critical density $3H^2M_{\rm Pl}^2$,
\be
\Omega_X\equiv\frac{\rho_X}{3H^2M^2_{\rm Pl}};\;\;\; \Omega_{\rm rad}\equiv\frac{\rho_{\rm rad}}{3H^2M^2_{\rm Pl}};\;\;\;
\Omega_{\rm kin,\varphi}\equiv \frac{\dot\varphi^2/2}{3H^2M^2_{\rm Pl}}=\frac{\varphi'^2}{6M^2_{\rm Pl}}\;,\;\;\;
\Omega_V\equiv \frac{V}{3H^2M^2_{\rm Pl}}\;,
\ee
where we use the 
and then the Friedmann equation becomes just the fact that the sum of all the $\Omega$'s is one,
\be
\Omega_X+\Omega_{\rm rad}+\Omega_{\rm kin,\varphi}+\Omega_V=1\;,
\ee
which implies
\be
\varphi'=M_{\rm Pl}\sqrt{6(1-\Omega_X-\Omega_{\rm rad}-\Omega_V)}.
\ee

The nontrivial equation is the KG equation, which becomes
\be
\varphi''+\left(\frac{3}{2}\Omega_X+\Omega_{\rm rad}+3\Omega_V\right)\varphi'=3cM_{\rm Pl}\Omega_X-\frac{1}{H^2}
\frac{dV}{d\varphi}\;,\label{KGnew}
\ee

We now have to solve this equation, given some initial conditions, which are a certain $\varphi=\varphi_i$, and an initial
``velocity" (derivative with respect to $N$) $\varphi'_i<0$. 
As suggested by the $V=0$ case considered in \cite{Hinterbichler:2013we}, we expect to find some attractor-like behaviour, where some of the energy density components scale in the same way\footnote{That is, we expect to find solutions in which the fractional densities $\Omega_i$ tend to a fixed-point behaviour.}. 

We must distinguish however between the cases of {\bf $c>0$}, when the coupling factor $F(\varphi)$ increases away from the
inflationary plateau, i.e. at small $\varphi$ (the plateau is at large $\varphi$), and the case of {\bf $c<0$}, when $F(\varphi)$ decreases in the
same direction as the potential, namely towards small $\varphi$.

In the $c>0$ case, we can have a local, time (or $N$) dependent  minimum of the effective potential $V_{\rm eff}$, which allows for the possibility
that $\rho_X$ and $\rho_V$ scale in the same way. Also the kinetic energy $\rho_{\rm kin,\varphi}$ can {\em a priori} scale the same way.

In the $c<0$ case, $\rho_V$ is approximately constant with $N$ on the plateau, but $\rho_X$ will decrease as $\sim e^{|c|\frac{\varphi}{M_{\rm Pl}}-3N}
\sim e^{-|c|\frac{|\varphi'|}{M_{\rm Pl}} N-3N}$, so it will be subleading. But the kinetic energy $\rho_{\rm kin,\varphi}$ can {\em a priori} scale in the same way.

In both cases, $\rho_{\rm rad}\sim e^{-4N}$, so it will subleading.

We will study the possible attractor behaviours in both cases, both analytically and numerically, in the next section.

\section{Evolution and attractors}

In order to find possible attractor-like behaviours, we consider an ansatz where the kinetic energy density $\rho_{\rm kin,
\varphi}$ is constant, so $\varphi'$ is constant,
\be
\varphi = \varphi_i - k N M_{\rm Pl}, \quad k> 0.\label{ansatz}
\ee
If moreover the Hubble constant $H$ is constant, as is expected if $\rho_V=V$ is approximately constant and the other terms are either
negligible or scale in the same way in (\ref{friedmann}), then also $\Omega_{\rm kin,\varphi}=\rho_{\rm kin,\varphi}/(3H^2M^2_{\rm Pl})$ is constant,
and so is $\Omega_V=V/(3H^2M^2_{\rm Pl})$, which is what we would mean by an attractor-like behaviour.  

Since as we saw, $\rho_{\rm rad}$ scales as $e^{-4N}$ so it quickly becomes subleading, for an attractor-like behaviour as above, we find
from the KG (\ref{KGnew}) and Friedmann (\ref{friedmann}) equations
\bea
3\left(\frac{1}{2}\Omega_X + \Omega_V\right)(-k) &=& 3c\Omega_X -\frac{1}{H^2}\frac{V'}{M_{\rm Pl}},\cr
\frac{k^2}{6} &=& 1 - \Omega_X - \Omega_V.\label{attractoreqs}
\eea

We now analyze separately the cases $c>0$ and $c<0$.

\subsection{The case $c>0$}

In this case, $\rho_X$ increases away from the plateau (at small $\varphi$), so we can have a local minimum of $V_{\rm eff}$.

{\bf Analytical results: Attractor 1}

It is clear from the attractor equations (\ref{attractoreqs}) that, in order to have an attractor, and if as we said we have $H\simeq$ constant,
we need to have $V'\equiv dV/d\varphi\simeq$ constant. 
Let us therefore assume that
\be
\frac{dV}{d\varphi}\equiv \a\equiv \tilde\a\frac{A}{M_{\rm Pl}}
\ee
is approximately constant, over a relevant number of e-folds $N$, and is
small in Planck units, i.e., $\a\ll M^3_{\rm Pl}$; moreover, $\tilde\a\ll 1$.
Here $A=\lambda M_{\rm Pl}^4$ is the plateau value for the potential. Then we have approximately
\be
\Omega_V=\frac{\rho_V}{3H^2M^2_{\rm Pl}}\simeq \frac{A}{3H^2M^2_{\rm Pl}}\equiv \Omega_{V,0}.
\ee
The definition of $\tilde \a$, together with the condition $\tilde \a\ll 1$ is done so that, in the context of inflation, the first slow-roll parameter
$\epsilon$ is small,
\be
\tilde \a=\sqrt{2\epsilon}\ll 1\Rightarrow \epsilon\ll 1.
\ee

Then, substituting $\Omega_X$ from the second equation in (\ref{attractoreqs}) into the first, and using the definition of $\tilde \a$ and
of $\Omega_V\simeq \Omega_{V,0}$, we find
\be
3\left(1-\Omega_V-\frac{k^2}{6}\right)\left(\frac{k}{2}+c\right)+3\Omega_V k=\frac{V'}{H^2M_{\rm Pl}}\simeq 3\tilde \a\Omega_V.\label{general}
\ee

Consider now an attractor with $\Omega_V$ scaling in the same way as $\Omega_X$, that is with
\be
\rho_X\propto e^{-c\frac{\varphi}{M_{\rm Pl}}-3N}\sim e^{(ck-3)N}\sim {\rm constant}\Rightarrow k=\frac{3}{c}.
\ee
Then we obtain
\be
3c\left[\left(1-\Omega_V-\frac{3}{2c^2}\right)\left(1+\frac{3}{2c^2}\right)+\frac{3}{c^2}\Omega_V\right]=\tilde \a\Omega_V.
\ee
Incidentally, note that the equations is invariant under changing simultaneously the signs of $c$ and $\tilde \a$. Indeed, this is what happens
if we redefine $\varphi$ to $\tilde \varphi=\varphi_i-\varphi$, to have a variable that increases from 0, instead of one that decreases from a  large
value: then in terms of $\tilde\varphi$, we change both the signs of $\tilde\a$ and of $c$. The more general equation (\ref{general}) is also
invariant under the simultaneous change of signs of $\tilde\a,c$ and $k$, for the same reason.

The above is a linear equation in $\Omega_V$, which is solved by
\be
\Omega_V\simeq \frac{1+\frac{3}{2c^2}}{1+\frac{\tilde\a/c}{1-\frac{3}{2c^2}}}.\label{approx}
\ee

We can now obtain some constraints on parameters. First, note that
\be
\Omega_X=1-\Omega_V-\frac{3}{2c^2}\geq 0\Rightarrow 1-\Omega_V\geq \frac{3}{2c^2} \Rightarrow \Omega_V\leq 1-\frac{3}{2c^2}.
\ee
Substituting the approximate value of $\Omega_V$ in (\ref{approx}), we obtain the condition
\be
\frac{\tilde\a}{c}=\frac{\tilde \a/c}{1-\frac{3}{2c^2}}\left(1-\frac{3}{2c^2}\right)\geq \frac{3}{c^2}\Rightarrow \tilde \a\geq \frac{3}{c}.
\ee

It seems then that, since we need $\tilde \a$ to be small we could have $c$ be very large, so that we still have $\tilde \a> 3/c$. The
problem is that then, we have $\Omega_V$ very close to 1. Expanding in $\tilde \a$, we get
\be
\Omega_V\simeq  1+\frac{3}{2c^2}-\frac{\tilde\a}{c}=1-\frac{1}{c}\left(\tilde\a-\frac{3}{2c}\right).
\ee

A set of reasonable values for the parameters $\tilde \a$ and $c$, that don't give too small values for $\Omega_X$ and $\Omega_{\rm kin,\varphi}$, are:
\be
(\sqrt{2\epsilon}=)\;\; \tilde \a\simeq \frac{1}{5}\;,\;\;\;\;c\sim 20.
\ee
These values satisfy the constraints above.
The resulting $\Omega$'s for them are
\be
\Omega_V\simeq 1-\frac{1}{100}\;,\;\;\;
\Omega_{\rm kin,\varphi}=\frac{3}{2c^2}\simeq \frac{3}{2\cdot 400}\simeq \frac{1}{300}\;,
\;\;\;
\Omega_X=1-\Omega_V-\Omega_{\rm kin}\simeq \frac{2}{300}.
\ee

These values are vey small, but measurable, so in this case we have a nontrivial attractor.

However, we need to remember that we will in fact compare with inflation in the next section, and that leaves a measurable imprint in the CMBR.
The correct analysis will be done in the next section, but for now we will assume that the parameters we find for the attractor-like solution are also the
parameters during inflation, which are measured in the CMBR. In this way, we will get oriented for the kind of values we need to take for $\tilde\a$
and $c$.

Since as we said, $\tilde\a=\sqrt{2\epsilon}$, assuming the $\epsilon$ here is the same relevant for the CMBR (which is not quite correct, as we will
see next section, but we will assume this for now), we must have
\be
n_s-1=1-6\epsilon+2\eta\simeq 0.97.
\ee
{\bf If $\eta\ll \epsilon$}, we have $\epsilon\simeq 1/200$, so $\tilde \a=\sqrt{2\epsilon}\simeq 1/10$. Then
$c>3/\tilde\a\sim 30$. Consider the value $c\sim 40$. Then we get the $\Omega$'s
\be
\Omega_V\simeq 1-\frac{1}{400}\;,\;\;\;
\Omega_{\rm kin,\varphi}=\frac{3}{2c^2}\simeq \frac{1}{1000}\;,\;\;\;
\Omega_X\simeq \frac{1}{700}\;,
\ee
which are still not unreasonable.

For the generic conformal inflation (\ref{generalpot}), we have
\be
\tilde \a=\left[\sqrt{\frac{2}{3}}B e^{-\sqrt{\frac{2}{3}}\frac{\varphi_i}{M_{\rm Pl}}}\right]e^{\sqrt{\frac{2}{3}}\frac{ (\varphi_i-\varphi)}{M_{\rm Pl}}}.\label{tildealpha}
\ee
From the above value $\tilde \a=1/10$, it means that the square bracket must equal this value. In the case where $B=2n$ (coming from a
power law function $g(x)$), this is indeed consistent with $\varphi_i\gg M_{\rm Pl}$.
Then
\be
\sqrt{\frac{2}{3}}\frac{\tilde \varphi}{M_{\rm Pl}}=\frac{\sqrt{6}}{c}N.
\ee
With $c\sim 40$, we get a coefficient of about 1/15. Then for $N\sim 5$ e-foldings, $\tilde \a$ varies by a factor
of $e^{1/3}\simeq 1.4$ (a 40$\%$ change), which is still reasonable, meaning that the attractor holds for about 5 e-foldings in this case.

In the case of the inverse power law potential (\ref{invpower}),
\be
\tilde \a=pB \left(\frac{\varphi}{M_{\rm Pl}}\right)^{-p-1}\;,
\ee
if we want the same $\tilde \a\sim 1/10$, for $p=2$ we obtain
\be
B\left(\frac{\varphi}{M_{\rm Pl}}\right)^{-3}\sim \frac{1}{20}\;,
\ee
and now we can choose an initial value $\varphi_i$ large enough so that the above quantity doesn't vary for many e-folds.

However, one more potential problem is that generically we have $|\eta|\gg \epsilon$, instead of $|\eta|\ll \epsilon$, as assumed until
now.

For the generic conformal inflation potential (\ref{generalpot}), we have
\be
\eta\equiv M^2_{\rm Pl}\frac{V''}{V}=-\frac{2}{3}Be^{-\sqrt{\frac{2}{3}}\frac{\varphi}{M_{\rm Pl}}}=-\sqrt{\frac{2}{3}}\tilde\a\;,
\ee
so indeed $|\eta|\gg \epsilon =\tilde \a^2/2$.

But then the scalar tilt is
\be
n_s-1\simeq 2\eta = -2\sqrt{\frac{2}{3}}\tilde \a\simeq \frac{3}{100}\;,
\ee
which implies that we must have
\be
\tilde \a \simeq \frac{1}{50}.
\ee
Since $c\geq 3/\tilde\a$, we  can choose for instance
\be
c\sim 200.
\ee
Then the ratios of energy densities are
\be
\Omega_V\simeq 1-\frac{1}{50\cdot 150}=1-\frac{1}{7500}\;,\;\;\;
\Omega_{\rm kin,\varphi}=\frac{3}{2c^2}=\frac{3}{8\cdot 10^4}\;,\;\;\;\;
\Omega_X=\frac{1}{7500}-\frac{3}{8\cdot 10^4}\sim \frac{1}{10^4}.
\ee
In this case,
\be
\frac{\varphi_i-\varphi}{M_{\rm Pl}}=\frac{3}{c}N\simeq \frac{N}{70}\;,\label{CIHattrac200}
\ee
which means that $\tilde \a$ and $\eta$ don't change significantly even over 70 e-folds, meaning that the attractor lasts at least as much.

For the inverse power law potential (\ref{invpower}),
\be
\eta =-p(p+1)B\left(\frac{\varphi}{M_{\rm Pl}}\right)^{-p-2}=-(p+1)\tilde \a\left(\frac{M_{\rm Pl}}{\varphi}\right)=-\frac{p+1}{p}\frac{2\epsilon}{B}
\left(\frac{\varphi}{M_{\rm Pl}}\right)^p.
\ee
Then generically, we also have $|\eta|\gg \epsilon$ (though with a sufficiently small $B$ and not too large $\varphi/M_{\rm Pl}$ we could avoid it).

But then, since $2\eta\simeq -3/100$, we get
\be
\tilde \a\frac{M_{\rm Pl}}{\varphi}\sim \frac{1}{100}\;,
\ee
which means, for $p=2$, that
\be
B\left(\frac{M_{\rm Pl}}{\varphi}\right)^4\sim\frac{1}{200}.\label{powerattrac200}
\ee
Then, with $B\sim 1$, we can impose $\varphi_i\sim 4 M_{\rm Pl}$,
and since as we saw, $\tilde \varphi/M_{\rm Pl}\simeq N/70$, $\tilde \a$ and $\eta$ are virtually unchanged even over 100 e-folds.

It is worth noticing that we didn't need to satisfy the CMBR constraints, since inflation is actually subsequent to the attractor-like era
considered in this section, as we will see in the next section. But we wanted to make sure that, even if we considered the tightest
constraints, we can still obtain a nontrivial attractor, so we used the CMBR constraints as if this attractor-like solution already describes inflation (which it does not). 

Finally, consider the fact that the effective potential
\be
V_{\rm eff}=V+\rho_X=V+\rho_m e^{-c\frac{(\varphi-\varphi_0)}{M_{\rm Pl}}}
\ee
has an instantaneous (at fixed $N$) minimum given by
\be
\frac{dV_{\rm eff}}{d\varphi}=0\Rightarrow \frac{\tilde \a A}{M_{\rm Pl}}\equiv V'(\varphi)=\frac{c}{M_{\rm Pl}}\rho_{m,0}e^{-c(\frac{\varphi_i-\varphi_0}{M_{\rm
Pl}}-kN)-3N}.\label{V'_{eff}}
\ee
But for the attractor, with $k=3/c$, the $N$ dependence cancels out on the right hand side of the above equation, which means that the
(late time) attractor sits (or rather, "tracks" it) at the instantaneous minimum of the effective potential, $\varphi_{\rm \rm att}(N)\simeq \varphi_{\rm min}(N)$,
meaning that there is a sort of adiabaticity.

{\bf Analytical results: Attractor 2}

However, there is still one more attractor-like case to consider, that will also play a role in the numerics.

In the previous case, we had assumed that $\rho_X$ scales like $\rho_V$, so we have a constant and nonzero $\Omega_X$.

If we consider instead that $\Omega_X=0$, just like $\Omega_{\rm rad}=0$ for the attractor, we obtain another one.
Then the Friedmann equation (\ref{friedmann}) becomes
\be
k\equiv -\frac{d\varphi}{M_{\rm Pl}dN}=\sqrt{6}\sqrt{1-\Omega_V}\;,
\ee
or equivalently
\be
\Omega_V=1-\frac{k^2}{6}.
\ee

On the other hand, the generic KG equation (\ref{KGnew}) in the assumption of an attractor ($d^2\varphi/dN^2=0$) with $\Omega_X=0$ becomes
\be
3\Omega_V M_{\rm Pl}k =\frac{V_0}{H^2}\frac{\tilde \a}{M_{\rm Pl}}\;,
\ee
which gives
\be
k\Omega_V=\frac{V_0}{3H^2M^2_{\rm Pl}}\tilde\a\Rightarrow k=\tilde \a.
\ee
For this attractor-like solution, we get
\be
\Omega_{\rm kin,\varphi}=\frac{\tilde \a^2}{6}\;,\;\;\;
\Omega_V=1-\frac{\tilde\a^2}{6}.\label{omegas}
\ee

Then, for the maximum value of $\tilde \a$ we considered (unrestricted by the CMBR), $\tilde \a\sim 1/5$, we get
\be
\Omega_{\rm kin,\varphi}\sim \frac{1}{150}\;,\;\;
\Omega_V\sim 1-\frac{1}{150}\;,
\ee
whereas for the CMBR-constrained values we obtain
\be
\Omega_{\rm kin,\varphi}\sim \frac{1}{1.5\times 10^4}\;,\;\;
\Omega_V\sim 1-\frac{1}{1.5\times 10^4}.
\ee

But of course, like for any dynamical system, if there is more than one attractor, each one of them has a ``basin of attraction", a well defined
region in the (phase) space of initial conditions, such that if we start inside it, we reach the given attractor.

To find initial conditions compatible with the $\Omega_X\neq 0$ attractor, we consider the fact that $H^2$ was scaled out of the KG equation near
the attractor, but was otherwise given by the Friedmann equation as
\be
H^2=\frac{\rho_V+\rho_{\rm rad}+\rho_X}{3M^2_{\rm Pl}(1-\varphi'^2/6)}=\frac{\rho_V/M^2_{\rm Pl}}{3(1-\varphi'^2/6)}(1+\Omega_X/\Omega_V).
\ee

Thus, since the overall $\rho_V$ just changes the scale of $H^2$,  if $\Omega_X/\Omega_V\simeq \Omega_{X,att.}/\Omega_{V,att}$, which
is about $10^{-4}$ in the $\tilde \a=1/50$ case, and if $\phi'\simeq \phi'_{\rm att.}=3/c$, we will be quickly driven to the $\Omega_X\neq 0$ attractor.

Otherwise, if for instance we start with $\Omega_X/\Omega_V\ll \Omega_{X, att.}/\Omega_{V, att.}$,
the competition between the two terms (with $\Omega_X$ and with $dV/d\phi$) on the right hand side of (\ref{KGnew}) means that
the other attractor will win.

Before turning to the numerical work, we note that we have defined the notion of ``attractor" in a physical sense, not in a strict mathematical
sense\footnote{We also used ``attractor-like solutions" to designate these physical ``attractors".  We let the search for a formal proof that these solutions are attractors in mathematical sense to further investigation.}. We defined it implicitly by having $d^2\varphi/dN^2\simeq 0$ (compared to other terms) for a long enough number of e-folds (certainly
larger than the number of e-folds needed to reach the ``attractor"). Also note that we cannot be sure there are no more more attractors (other than
the two above), but our extensive numerical work seems to suggest that there aren't.

{\bf Numerical results: Attractor 1}

\begin{figure}
\centering
\includegraphics[width=0.5\linewidth]{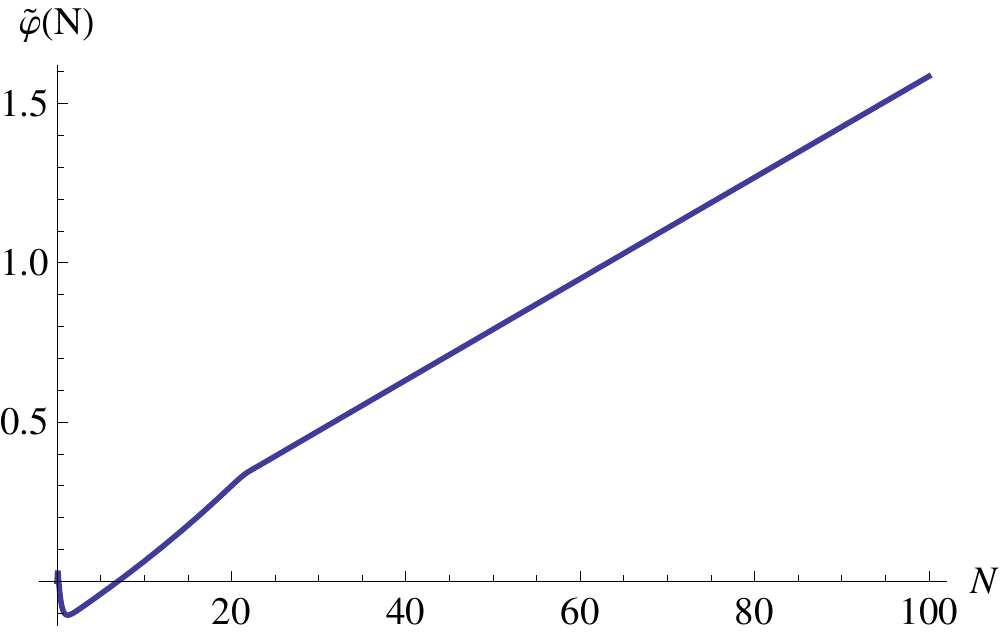}
\caption{Conformal inflation attractor $\tilde{\varphi}(N) = \varphi_i -\varphi(N) $ with $\varphi_0 = \varphi_i =
\sqrt{\frac{3}{2}}\ln(\sqrt{\frac{2}{3}} \frac{B}{\tilde{\alpha}})$, $\varphi_i' = -(\frac{3}{c}+0.5)$. We have $\varphi(N)
 = -0.0158977 N \approx -3N/c$ once the attractor is reached.}
\label{fig:CIH_attractor_c200}
\end{figure}

\begin{figure}
\centering
\includegraphics[width=0.5\linewidth]{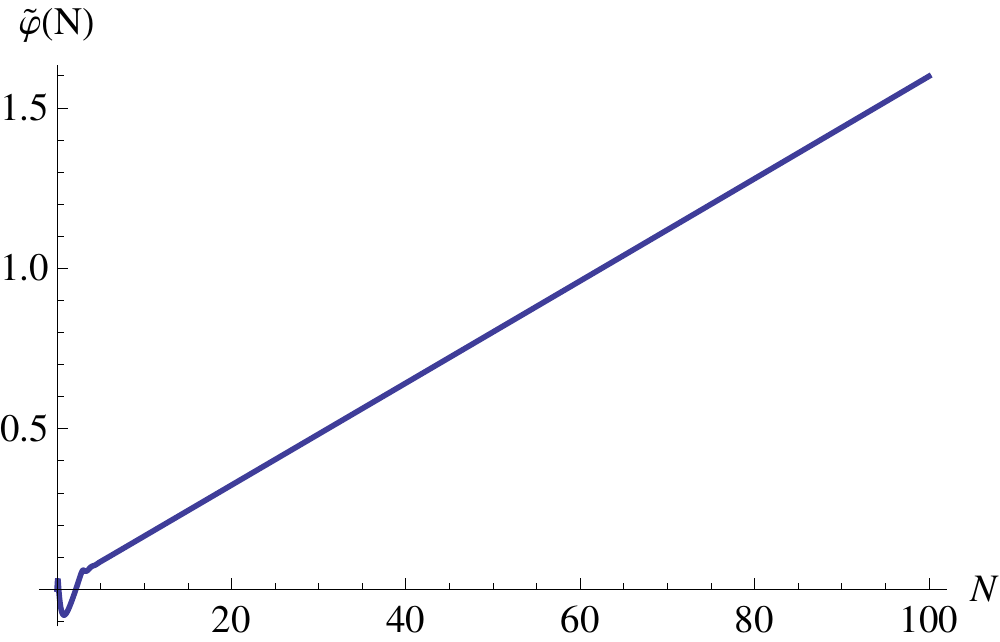}
\caption{Inverse power law attractor $\tilde{\varphi}(N) = \varphi_i -\varphi(N) $ with $\varphi_0= \varphi_i = 3 $,
$\varphi_i' = -(\frac{3}{c}+0.5)$. In this case, $\varphi(N) = -0.0159053 N\approx - 3N/c$ after onset of the attractor.}
\label{fig:powerlaw_attractor_c200_2}
\end{figure}

We solved numerically the KG equation (\ref{KGnew}) for $\varphi(N)$ with $H$ given by (\ref{friedmann}), ensuring a flat universe.
For generic conformal inflation, we used 
$B= 6$ while for the inverse power law potential $B = 1$ and $p=2$, with $A=10^{-9}$, $\rho_{m,0} = 10^{-4}A$, $\rho_{rad,0}=0.5 A$,
$\tilde{\alpha} = 1/50$ and $c= 3/\tilde{\alpha}+40$ in both cases (we used Planckian units in all numerical simulations). We considered
the attractors-like solutions (\ref{CIHattrac200}) and (\ref{powerattrac200}), since they last enough to have sufficient number of e-folds of inflation.
The numerical solution corresponding to them are shown in Figure \ref{fig:CIH_attractor_c200} and Figure \ref{fig:powerlaw_attractor_c200_2}.


\begin{figure}
\centering
\includegraphics[width=0.5\linewidth]{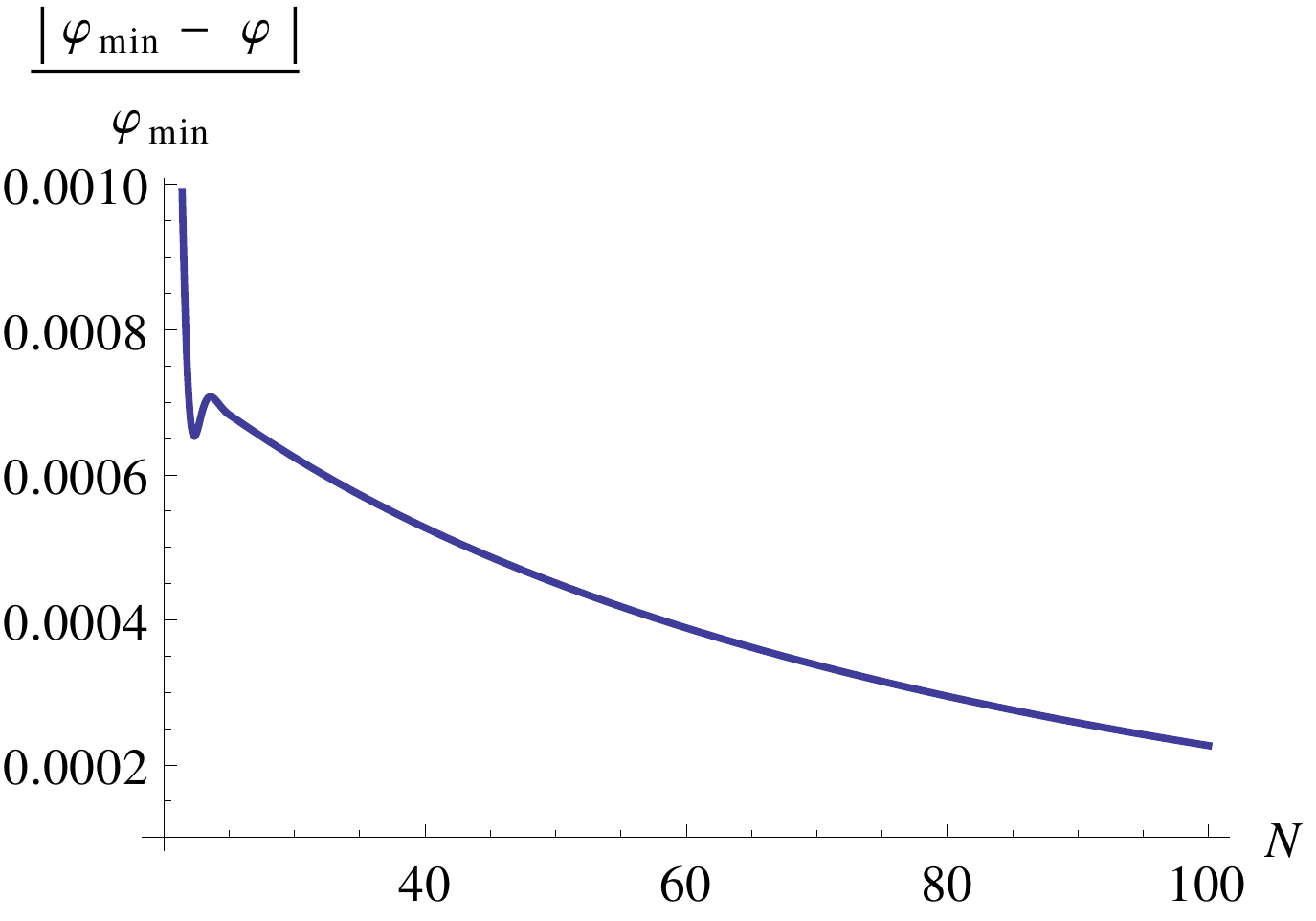}
\caption{Comparison between the instantaneous minimum of effective potential and numerical (attractor) solution.}
\label{fig:CIH_phimin_phiattrac}
\end{figure}

For generic conformal inflation, we can calculate the exact minimum of the potential at fixed $N$ and compare with
the numerical solution, to verify if the attractor follow the instantaneous minimum of the effective potential. This is
shown in Figure \ref{fig:CIH_phimin_phiattrac}, from which we found numerically that at late times the attractor solution sits at the minimum.

\begin{figure}
\centering
\includegraphics[width=0.5\linewidth]{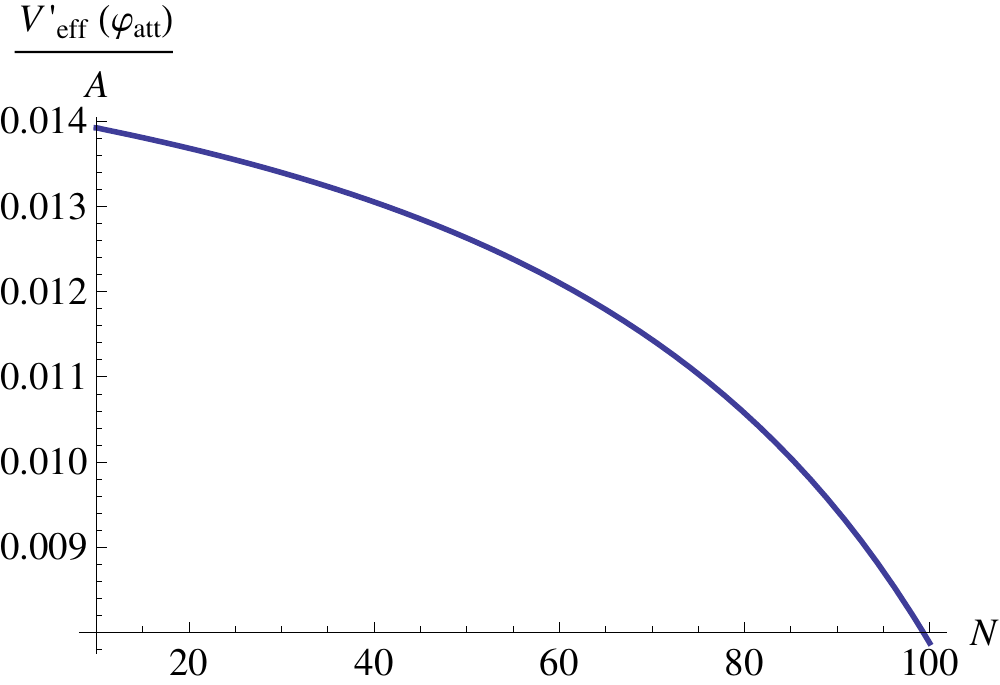}
\caption{The derivative of the effective potential using the inverse power law potential is effectively zero after $N=25$ e-folds.
The numerical solution follows the minimum of the effective potential. $A$ is the amplitude of the potential.}
\label{attractor_follow_minimum_Powerlawpotential}
\end{figure}

For the  inverse power law potential, the instantaneous minimum satisfies a transcendental equation and in order to check if the
attractor follows it, we computed $V'_{\rm eff}(\varphi(N),N)$ at numerical solution.  If $\varphi_{\rm att}(N) \simeq \varphi_{\rm min}(N)$,
we should have $V'_{\rm eff}(\varphi_{\rm att}(N))\simeq 0$, after the attractor is reached. Figure \ref{attractor_follow_minimum_Powerlawpotential}
shows that this is actually the case.


{\bf Numerical results: Attractor 2}

In this case, we expect the attractor-like solution to last less than attractor $1$. This is so because $k = \tilde{\alpha}$ and then from $c>3/\tilde{\alpha}$
we have that $\phi(N)$ changes faster with $N$. Also, since $\Omega_X \sim e^{-c\varphi-3N}\sim e^{-(3/\tilde{\alpha} + \Delta c)(-\tilde{\alpha}N)
-3N}\sim e^{\tilde{\alpha}\Delta c N}$, we see that even starting with very small matter density it will increase so that  numerical solutions will
only approximate attractor 2.

\begin{figure}
\centering
\includegraphics[width=0.5\linewidth]{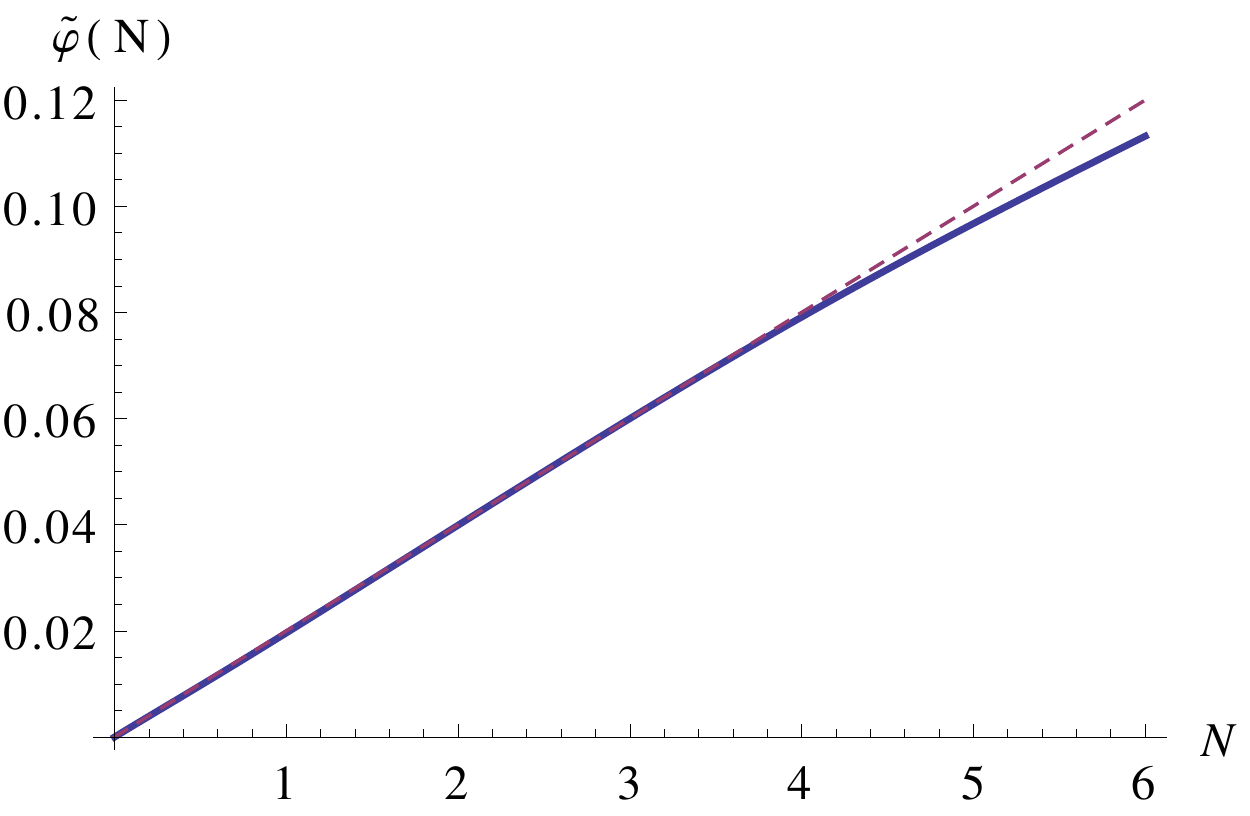}
\caption{Numerical solution (thick line) for conformal inflation potential and analytical result (dashed line) for the attractor 2.}
\label{fig:CIH_attractor2}
\end{figure}

\begin{figure}
\centering
\includegraphics[width=0.5\linewidth]{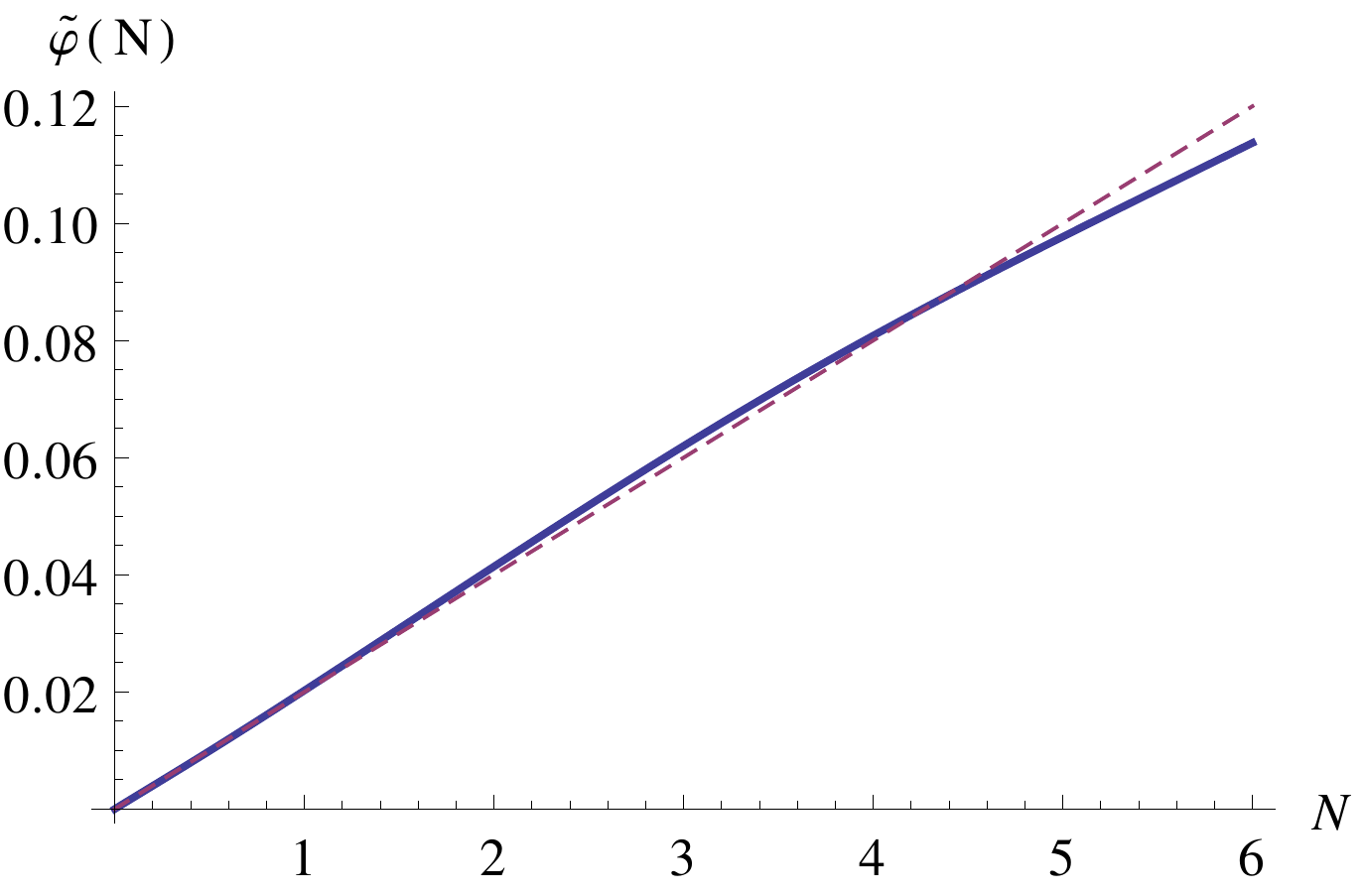}
\caption{Numerical solution (thick line) for inverse power law potential and analytical result (dashed line) for the attractor 2.}
\label{fig:powerlaw_attractor2}
\end{figure}

We consider the $\tilde{\alpha} = 1/50$ case with $\rho_{X,0} = 10^{-6}A$, such that $\Omega_{X,0}/\Omega_{V,0}\ll 10^{-4}$. Figures
\ref{fig:CIH_attractor2} and \ref{fig:powerlaw_attractor2} show the numerical solutions for generic conformal inflation and inverse power
law potentials. In the inverse power law case $\varphi_i = \varphi_0 = 4.64$, and $\varphi'_i= -\tilde{\alpha}$ for both potentials. All other
parameters are the same as attractor $1$. We see that attractor $2$ doesn't last more than $5$ e-folds.


\subsection{The case $c<0$}

In this case, $\rho_X$ decreases in the same direction, towards small $\varphi$, so there is no local minimum for $V_{\rm eff}$.

{\bf Analytical results}

In this case, the second attractor from the previous subsection ($c>0$ case) is still valid. Indeed, we saw that this attractor-like solution corresponds
to $\Omega_X=0$, so in this case the sign of the exponent of the coupling in $\Omega_X$ is irrelevant.

Since moreover, in this case, there cannot be any nontrivial attractor with $\Omega_X\neq 0$, since $\Omega_X
\sim e^{-c\frac{\varphi}{M_{\rm Pl}}-3N}\sim e^{-|c|\frac{|\varphi'|}{M_{\rm Pl}}N-3N}$ decays, while $\Omega_V$ stays constant, the only possible
endpoint for any initial condition is the attractor with $\Omega_X=0$. We will call this attractor a {\em kinetic-potential phase}.

Note that it is different from the case of inflation, in which $\Omega_V$ dominates, and the kinetic energy is generically small, and
then starts increasing.

The nonzero energy densities are still given by (\ref{omegas}), but this time we don't have any constraints on the value of $\tilde\a$
(like $\tilde \a\geq 3/c$ and $\Omega_V\simeq 1+3/(2c^2)-\tilde\a/c$  before), so we can fix it as we like. However, we still find that,
for the value $\tilde \a\sim 1/50$, consistent with CMBR in the hypothesis of inflation, the attractor is maintained for over 70 e-folds.

Generic initial conditions should lead to the attractor, but we must ensure that the attractor is reached in less than the number of e-folds it persists.

An interesting case, however, is now possible: we can have a kind of {\em kinetic domination}, usually called {\em kination}. Not quite that,
of course, since in fact this is an attractor with small, though constant and nonzero, $\Omega_V$. That is, we can have
$k=\tilde\a\simeq \sqrt{6}$, so that
\be
\Omega_{\rm kin,\varphi}=\frac{\varphi'^2}{6M^2_{\rm Pl}}\simeq 1
\ee
and $\Omega_V$ very small, though nonzero, whereas $\Omega_{\rm rad}$ and $\Omega_X$ are truly 0.

However, we still expect to find a condition on $c$ for the existence of this attractor-like solution, since at $c=0$ (no chameleon), we have no
attractor (this is the standard inflationary case, which doesn't admit the kination phase).
To understand this qualitatively, we rewrite the KG equation (\ref{KGnew}), in the case
$\Omega_{\rm rad}=0$, as
\be
H^2\varphi''+\left(\rho_V+\frac{\rho_X}{2}\right)
\frac{1}{M^2_{\rm Pl}}\varphi'=c\frac{\rho_X}{M_{\rm Pl}}-\frac{dV}{d\varphi}.
\ee
This takes the form of a ``force law", with the first term on the left hand side being the ``mass times acceleration", and the second being a
friction term, proportional to the velocity and opposing the acceleration.  Then the term $-dV/ d\varphi<0$ is a force driving us towards
smaller $\varphi$, and the first term, for $c<0$, is another driving force, in the same direction as the potential, allowing the constant
velocity motion to go on for longer. If the initial condition has a very large $\rho_X$, this term will dominate initially, so even if after a long time
it decays to zero, its effect is felt through the settling the initial motion into the attractor.
This suggests that there should be some minimum value for $c$, depending on the initial conditions, below which we don't get the attractor-like behaviour.

{\bf Numerical results}

We considered just the generic conformal inflation potential for this case.

For $\rho_X = 0$, we are back to the usual inflationary scenario. In this case, the
conformal inflation potential is a plateau for large field values and becomes steep right before becoming negative. For the parameters
used in previous section we have $\epsilon \simeq 0.21$ for $\varphi = 3$, so inflation ends after around that. For typical initial conditions,
with $\varphi_i$ well at the plateau and initial velocity smaller than the critical value $\sqrt{6}$, the field roll the potential slowly.

\begin{figure}
\centering
\includegraphics[width=0.5\linewidth]{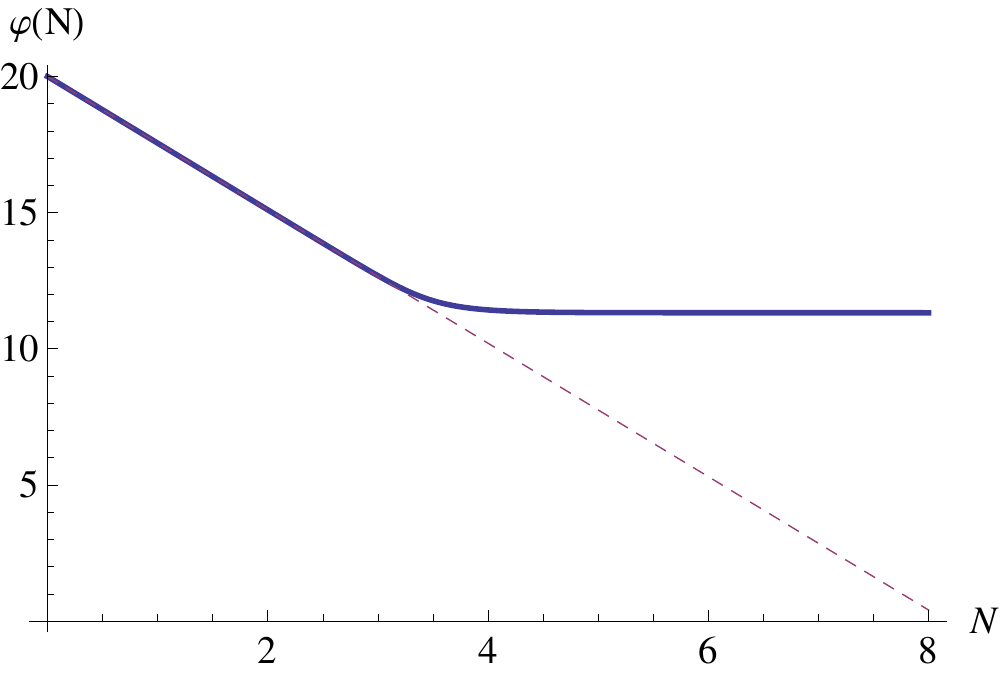}
\caption{Numerical solution with $\rho_{X,0} = e^{10^{-2}}$, $\varphi_i = \varphi_0 = 20$, $\varphi'_i = -1$, $\rho_{r,0} = 0.5 A$,
$c = -200$ (thick line) and $c=-571$ (dashed). All other parameters are the same as before. For these initial conditions,
$c_{\text{min}} = -571$ and the dashed line corresponds to $\varphi(N) = \varphi_i- \sqrt{6}N$.}
\label{fig:c_negative2}
\end{figure}

This picture changes with $\rho_X \neq 0$, where we can have the kinetic dominated attractor. But if $|c|$ is not big enough, we found numerically
that kination phase do not last much and $\varphi$ starts to roll down the potential slowly until inflation end around $\varphi \simeq 3$.
But increasing the value of $|c|$ we get kination, as show in Figure \ref{fig:c_negative2}.

Table ($1$) shows what is the minimal value of $|c|$ in terms of $\rho_{X,0}$ such that the kinetic dominated attractor-like behaviour is achieved.
Other parameters are the same used previously.

\begin{table}[h!]
	\centering
	\label{table}
	\begin{tabular}{ccc}
		\toprule
		$\rho_{X,0}$ & $c_{\text{min}}$ &\\
		\toprule
		$e^{10^{-2}}$ & $-571$ & \\
		\toprule
		$e^{10^{-1}}$ & $-186$ & \\
		\toprule
		$1$ & $-179$ & \\
		\toprule
		$e^{10}$ & $-70$ & \\
		\toprule
		$e^{10^2}$ & $-2$ & \\
		\bottomrule
	\end{tabular}
	\caption{Minimal value of $c$ from which we get kination for specific values
		of $\rho_X$.}
\end{table}

\section{Modifications to inflationary era and CMBR observables}

In this section, we explore the consequences  for inflation of the existence of the attractor-like behaviours found in the previous section.

\subsection{The case $c<0$ and shortened inflation}

In this case, the chameleon coupling term decreases in the same direction as the potential, and we saw that nevertheless we found a kinetic-potential
attractor, with a constant ``velocity" $d\varphi/dN$ and a constant $\Omega_V$. More interestingly, we could have an {\em almost kination}
attractor, when $\Omega_{\rm kin,\varphi}\simeq  1$.
Normally, in inflation the scalar rolls down the potential slowly, and then accelerates as the slope steepens, finally ending inflation.

But in the current case, the main effect of the chameleon coupling is to start with a  kinetic (almost kination) phase, and thus delay the onset of the
region of inflation {\em per se}. Once the attractor behaviour ends, and the $\rho_X$ component has sufficiently decayed so as to become irrelevant,
we are back to usual inflation. This of course depends on the initial value for $\rho_X$ at the start of the kinetic  phase.

\begin{figure}
\centering
\includegraphics[width=0.5\linewidth]{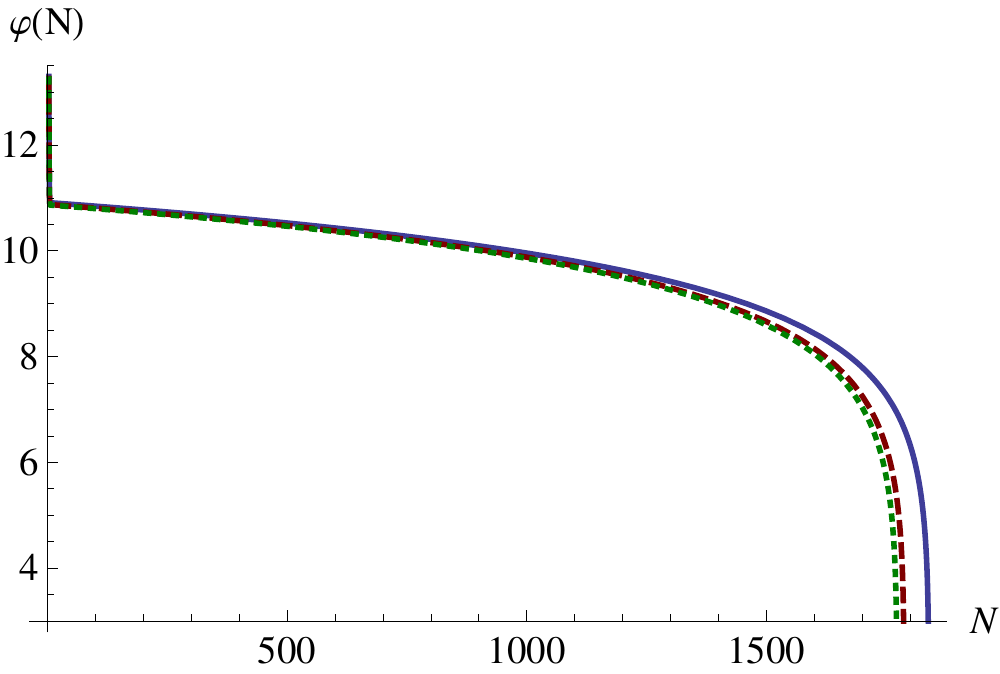}
\caption{End of inflation in numerical solutions with $\rho_{X,0} = e^{10^{-2}}$ (blue thick), $\rho_{X,0}=e^{10^{-1}}$ (red dashed) and $\rho_{X,0}= e^{10}$ (green dotted). In all these cases, the kination phase ends before $N= 5$.}
\label{fig:c_negative3}
\end{figure}

To compare with pure inflation case ($c = 0$) we need to set the initial value of $\varphi$ such that transition from kinetic phase to normal inflation occurs before $\varphi \simeq 3$. But then pure inflation will remain for a large number of e-folds.
Indeed, using the parameter values as in Figure \ref{fig:c_negative2}, we get that inflation ends around $N = 1.7\times 10^6$. Now, for $c\neq0$ and varying $\rho_{X,0}$, we found numerically that inflation ends earlier than that, assuming that it ends when $\varphi$ reaches $3$. Figure \ref{fig:c_negative3} shows numerical solutions with $c = -50$ and different values of $\rho_{X,0}$.

\subsection{The case $c>0$ and modified inflation}

In this case, the chameleon coupling term increases in the direction that the potential decreases, leading to the existence of an instantaneous
(at fixed $N$) minimum of the effective potential $V_{\rm eff}=V+\rho_X$. As we saw, the attractor 1 (with $\Omega_X\neq 0$) follows at
late times this instantaneous minimum $\varphi_{\rm min}(N)$ of the effective potential, $\varphi_{\rm att}(N)=\varphi_{\rm min}(N)$.

Unlike the $c<0$ case then, we now have a motion that is qualitatively modified: the motion is not due to rolling down the potential itself anymore, but
rather to the variation with $N$ of the instantaneous minimum of $V_{\rm eff}$, where the scalar field sits. It is also not clear how and when this
behaviour will end. We will look more into it in the next subsection. 

But the result is that we have a new form of inflation, a {\em modified inflation phase}. Then strictly speaking, we have to calculate the
spectrum of perturbations in this phase ab initio. In order to do so we rewrite our model in terms of a second scalar field that has a non-minimal kinetic coupling with the chameleon field. We show that such a model leads to equations of motion that are equivalent to the ones we had studied in the previous sections. The perturbations for two scalar fields where one of them has a non-minimal kinetic coupling and some general potential were computed in \cite{DiMarco:2002eb}. We will apply their results to our scenario. 

Another important consistent check for the model is whether the presence of a second scalar field will lead to an increase of the entropy fluctuations. In the following section we check that the chameleon coupling in our model actually prevents the entropy modes to grow during inflation. We are then allowed to compute the scalar tilt and tensor to scalar ratio parameters in terms of the adiabatic modes only.

Unfortunately, in trying to compare with CMBR data, we face a quandary: we saw that the attractor behaviour $\varphi_{\rm att}(N)$ seems to
go on forever, and there is no way to end inflation. That is relevant, since the scale that we see the CMBR in the sky now, $k_0\sim 10^{-3} Mpc^{-1}$,
is situated a number $N=N(k_0)$ e-folds of inflation before the {\em end of inflation}, related (by the known evolution of the Universe,
assuming a normal Einstein-gravity radiation dominated phase after reheating until Big Bang Nucleonsynthesis) by the usual formula
\be
N=\ln \frac{a_{\rm end}}{a_{k_0}}=56 -\frac{2}{3}\ln \frac{10^{16}GeV}{\rho_*^{1/4}}-\frac{1}{3}\ln \frac{10^9 GeV}{T_R}\;,
\ee
where $\rho_*$ is the energy density at the end of inflation and $T_R$ is the reheat temperature.

Assuming some standard values for $\rho_*$ and $T_R$, one gets around 60 e-folds of inflation from the scale $k_0$ (relevant for CMBR)
exiting the horizon, when we should measure $n_s$ and $r$, and the end of inflation.

Thus in order to be consistent with observations, we must find a way to end inflation.

\subsection{Ending inflation for $c>0$}

The simplest way to end inflation in the case $c>0$ is to remember that the heavy non-relativistic particle(s) that make up $\rho_X$ must decay,
since they are not there after inflation. That is, they must have a decay constant $\Gamma=1/\tau$ with $\tau$ of the order of the age of the
Universe at the end of inflation. Then the energy density $\rho_m$ of this component satisfies the usual modified equation of motion
\be
\dot \rho_m +3H \rho_m +\Gamma\rho_m=0.
\ee

This implies an extra factor in the decreases of $\rho_X=\rho_m F(\varphi)$ with $N$ of
\be
e^{-\Gamma t}=e^{-\frac{\Gamma}{H}Ht}\sim a^{-\frac{\Gamma}{H}}=e^{-\frac{\Gamma}{H}N}.
\ee
Then, for instance for a $\Gamma\sim 0.1 H$ (or $\tau\sim 10 H^{-1}$), we would obtain an extra factor of $e^{-0.1N}$. After about 100 e-folds,
this would give a contribution of about $e^{-10}\simeq 10^{-4.3}$.

However, the drawback of this method is that it goes on very slowly, instead of the sudden end to inflation that we usually need.

One alternative possibility  would be to say that, once the
slope of the potential $V(\varphi)$ starts becoming steep (so that its own $\epsilon$ or $\eta$ would be of order 1, so that normal inflation
would have ended), some unknown particle physics mechanism would allow $\rho_m$ to decay {\em into the inflaton/chameleon particles
themselves},  thus ending inflation, and allowing reheating (the conversion of the inflaton into lighter particles) to happen.

This can be implemented as an ad-hoc turning off of $\rho_X$ at the same time normal inflation would have ended, though
it is not clear how one could implement it in detail from a particle physics perspective. In the rest of this section we 
calculate the CMBR observables as a function of the number of e-folds of inflation.


\subsection{Microscopic description}

To proceed with the calculation of the inflationary observables, we show the consistency of our model with the one considered in \cite{DiMarco:2002eb}. In fact, the equations of motion we have assumed during this paper have a microscopic description given by the Lagrangian:
\begin{align}
    S= \int d^4x \sqrt{-g}\left[\frac{M_{Pl}^2}{2}R - \frac{1}{2}g^{\mu\nu}\partial_\mu\varphi\partial_\nu\varphi - V(\varphi)\right] + S_m[\tilde{g}_{\mu\nu} = F^2(\varphi)g_{\mu\nu}],
\end{align}
where the non-relativistic matter can be described, without loss of generality,  by a scalar field, 
\begin{align}
    S_m = \int d^4x \sqrt{-\tilde{g}}\left[-\frac{1}{2}\tilde{g}^{\mu\nu}\partial_\mu\chi\partial_\nu\chi - U(\chi)\right].
\end{align}
In terms of the Einstein-frame metric the above action can be rewritten as 
\begin{align}\label{robert}
  S =   \int d^4x \sqrt{-g}\left[\frac{M_{Pl}^2}{2}R - \frac{1}{2}g^{\mu\nu}\partial_\mu\varphi\partial_\nu\varphi - \frac{1}{2}F^2(\varphi) g^{\mu\nu}\partial_\mu\chi\partial_\nu\chi - \tilde{V}(\varphi, \chi)\right],
\end{align}
which has exactly the form considered in \cite{DiMarco:2002eb} with 
\begin{align}
  F(\varphi) = e^{b(\varphi)}, \quad  b(\varphi)= -c(\varphi-\varphi_0)/M_{pl},  \quad \tilde{V}(\varphi, \chi) = V(\varphi)+ F^4(\varphi)U(\chi).
\end{align}

In our case we consider everything in terms of the inflaton and the energy density of a matter ($p=0$ in the Einstein-frame) fluid. We need to find a mapping between the energy density\footnote{Remember that $\rho_X = F(\varphi)\rho_m$.} $\rho_m$ and the second scalar field $\chi$. In order to do so, we compute 
\begin{align}
    \tilde{T}_{\mu\nu} = -\frac{2}{\sqrt{-\tilde{g}}}\frac{\delta S}{\delta \tilde{g}^{\mu\nu}}
\end{align}
and then use\footnote{See \cite{Waterhouse:2006wv} for a detailed proof of this relation.} $\rho_m = F(\varphi)^3 \tilde{\rho}_m$. Assuming a homogeneous field, we have that
\begin{align}
    \tilde{T}_{\mu\nu} = \partial_\mu\chi\partial_\nu\chi - \tilde{g}_{\mu\nu}\left(\frac{1}{2}\tilde{g}^{\mu\nu}\partial_\mu\chi\partial_\nu\chi + U\right)\nonumber\\
    \implies \tilde{T}_{00} = \frac{1}{2}\Dot{\chi}^2 + F^2U \implies \tilde{\rho}_m = -\tilde{T}^0_0 = \frac{1}{2}F^{-2}\Dot{\chi}^2 + U.
\end{align}
Then we find
\begin{align}
\rho_m = \frac{1}{2}F\Dot{\chi}^2+ F^3U.
\end{align}

Also, from $T_{\mu\nu} = F^2\tilde{T}_{\mu\nu}$ and
\begin{align}
    \tilde{T}_{ij} = \tilde{g}_{ij}F^{-2}(\frac{1}{2}\Dot{\chi}^2- F^2 U)
\end{align}
we have
\begin{align}
    T_{ij} = p g_{ij} = F^2\tilde{p}\tilde{g}_{ij} \implies p = F^4\tilde{p} = F^2 (\frac{1}{2}\Dot{\chi}^2- F^2 U).
\end{align}
Therefore, the condition for $\chi$ to describe a matter fluid, $p=0$, implies $\frac{1}{2}\Dot{\chi}^2 = F^2U$. Using this into the expression for $\rho_m$, we find
\begin{align}
\rho_m = F\Dot{\chi}^2.
\end{align}

From the above mapping, one can rewrite the equation of motion coming from action (\ref{robert})
\begin{align}
    \Ddot{\varphi} + 3H \Dot{\varphi} + \frac{\partial\tilde{V}}{\partial\varphi} &= \frac{\partial b}{\partial \varphi} e^{2b}\Dot{\chi}^2,\nonumber\\
    \Ddot{\chi} + (3H + 2 \frac{\partial b}{\partial\varphi}\Dot{\varphi})\Dot{\chi} + e^{-2b}\frac{\partial \tilde{V}}{\partial \chi} &= 0,
\end{align}
as 
\begin{align}
    \Ddot{\varphi} + 3H \Dot{\varphi} + \frac{\partial V}{\partial\varphi} &= -\rho_m \frac{\partial  F}{\partial\varphi},
    \nonumber \\
    \Dot{\rho}_m + 3H\rho_m &= 0,
\end{align}
which are exactly the equations we have used throughout this paper. Note that to find it we had to use the matter 
fluid condition, since the chameleon coupling is sensitive to the equation of state of the fluid.

Adiabatic and entropy perturbations for the action in (\ref{robert}) were also computed in detail in \cite{DiMarco:2002eb}. Having showed our macroscopic description is recovered by the microscopic one given by (\ref{robert}) we now move on to investigate whether these perturbations are under control in our scenario.

\subsubsection{Adiabatic and Entropy Perturbations}

Following \cite{DiMarco:2002eb}, in order to proceed with calculations we first decompose the fields in its adiabatic and entropy components (see also \cite{Langlois:2008mn, Lalak:2007vi, Gordon:2000hv}):
\begin{align}
    d\sigma &= \cos\theta d\varphi + e^b \sin\theta d\chi\nonumber\\
    d s&= e^b\cos\theta d\chi - \sin\theta d\varphi,
\end{align}
where
\begin{align}
    \cos\theta = \frac{\Dot{\varphi}}{\Dot{\sigma}}, \quad \sin\theta = \frac{e^b\Dot{\chi}}{\Dot{\sigma}}, \quad \text{with} \quad \Dot{\sigma} = \sqrt{\Dot{\varphi}^2 + e^{2b}\Dot{\chi}^2},
\end{align}
and we write the metric perturbations in longitudinal gauge
\begin{align}
    ds^2 = -(1+2\Phi)dt^2 + a^2(t)(1-2\Phi)\delta_{ij}dx^i dx^j.
\end{align}

The curvature and entropy perturbations, $\zeta$ and $\delta s$, defined as
\begin{align}
\zeta = \Phi - \frac{H}{\Dot{H}}(\Dot{\Phi}+ H\Phi)\nonumber\\
\delta s = -\frac{\Dot{\sigma}}{2H(d\tilde{V}/ds)}(\Dot{\zeta} - \frac{k^2}{a^2}\frac{H}{\Dot{H}}\Phi)
\end{align}
satisfy the following equations of motion
\begin{align}\label{adiabatic}
    \Ddot{\zeta}+ \left(3H - 2\frac{\Dot{H}}{H}+ \frac{\Ddot{H}}{\Dot{H}}\right)\Dot{\zeta} + \frac{k^2}{a^2}\zeta = \frac{H}{\Dot{\sigma}}&\left[\frac{d}{dt}(\Dot{\theta}\delta s)-2 \left(\frac{1}{\Dot{\sigma}}\frac{d \tilde{V}}{d\sigma}+ \frac{\Dot{H}}{H}\right)\Dot{\theta}\delta s\right. \nonumber\\
    &\left.+ 2 \frac{db}{d\varphi}h(t) + \frac{d^2b}{d\varphi^2} \Dot{\sigma}^2 \sin 2\theta \delta s\right]
\end{align}
\begin{align}\label{entropy}
    \delta \Ddot{s} + 3H\delta \Dot{s} + &\left[\frac{k^2}{a^2}+ \frac{d^2 \tilde{V}}{ds^2}+ 3\Dot{\theta}^2+ \left(\frac{d b}{d\varphi}\right)^2g(t)+\frac{d b}{d\varphi}f(t)- \right.\nonumber\\
    &\left.- \frac{d^2 b}{d\varphi^2}\Dot{\sigma}^2- 4\frac{1}{\Dot{\sigma}^2}\left(\frac{d \tilde{V}}{ds}\right)^2\right]\delta s = 2\frac{1}{H}\frac{d \tilde{V}}{ds}\Dot{\zeta},
\end{align}
 where
\begin{align}
    f(t)&= \frac{d\tilde{V}}{d\varphi}(1+ \sin^2\theta) - 4\frac{d\tilde{V}}{ds}\sin\theta\nonumber\\
    g(t)&= -\Dot{\sigma}^2(1+3\sin^2\theta)\nonumber\\
    h(t)& = \Dot{\sigma}\frac{d}{dt}(\sin\theta \delta s)-\sin\theta\left(\frac{\Dot{H}}{H}\Dot{\sigma}+ 2\frac{d \tilde{V}}{d\sigma}\right)\delta s - 3H \Dot{\sigma}\sin\theta \delta s\nonumber\nonumber\\
    \Dot{\theta} &= \Dot{\sigma}\left(-\frac{1}{\Dot{\sigma}^2}\frac{d \tilde{V}}{d s}- \frac{db}{d\varphi}\sin\theta\right).
\end{align}
In order to apply the above results to our model we only need to write all functions in terms of $N$ and to replace all possible $\chi$ dependence for $\rho$.

\subsubsection{Perturbations in conformal inflation with chameleon coupling}

From equation (\ref{adiabatic}) and (\ref{entropy}), we see that entropy perturbations will feed adiabatic modes and vice versa. Since there is no evidence for entropy modes in the CMB data, we must check if entropy perturbations are under control in our model. In order to do so, we rewrite equation (\ref{entropy}) as
\begin{align}
   \delta \Ddot{s} + 3H\delta \Dot{s} + \left(\frac{k^2}{a^2}+ m_{\text{eff}}^2\right)\delta s = -\frac{k^2}{a^2}\frac{d \tilde{V}}{ds}\frac{4M_{Pl}^2\Phi}{\Dot{\sigma}^2},\nonumber\\
    m_{\text{eff}}^2 \equiv \frac{d^2 \tilde{V}}{ds^2}+ 3\Dot{\theta}^2+ \left(\frac{d b}{d\varphi}\right)^2g(t)+\frac{d b}{d\varphi}f(t)- \frac{d^2 b}{d\varphi^2}\Dot{\sigma}^2.
\end{align}
One can see that the entropy evolves freely at super-Hubble scales. Moreover, in order for small-scale quantum fluctuations to generate super-Hubble perturbations during inflation, the fluctuation should be light compared to the Hubble 
scale. In fact, the existence of super-Hubble entropy perturbation requires that
\begin{align}
    m_{\text{eff}}^2 < \frac{9}{4}H^2,
\end{align}
otherwise the perturbations stay at the vacuum state and there is a strong suppression at large scales. 

\begin{figure}
\centering
\includegraphics[width=0.5\linewidth]{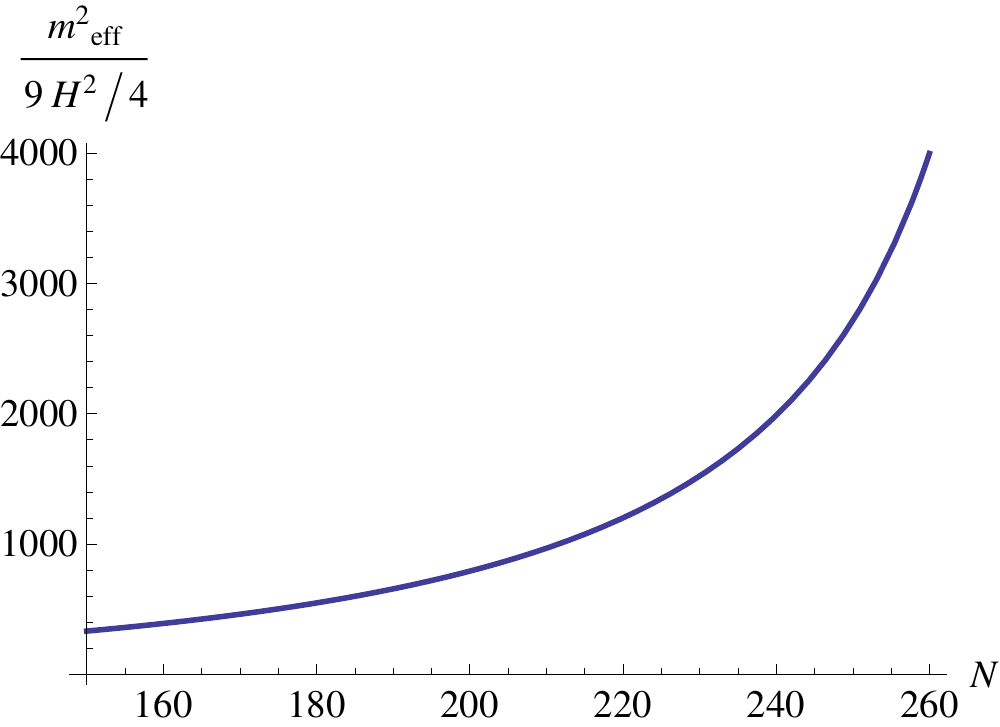}
\caption{Evolution of the effective mass for the \textit{conformal inflation model} during inflation, evaluated for the solution showed in Figure \ref{fig:CIH_attractor_c200}. Entropy perturbation field gets heavier due to the coupling to the chameleon field.}
\label{ci}
\end{figure}

\begin{figure}
\centering
\includegraphics[width=0.5\linewidth]{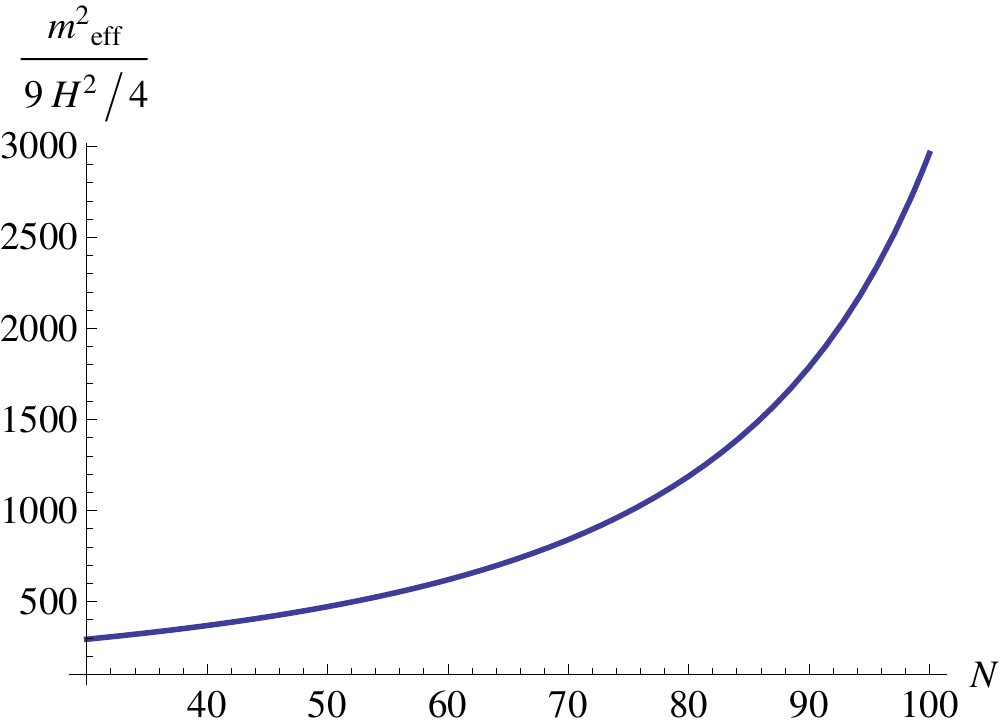}
\caption{Evolution of the effective mass for the \textit{power law model} during inflation, evaluated for the solution displayed at Figure \ref{fig:powerlaw_attractor_c200_2}. Entropy perturbation field also gets heavier due to the coupling to the chameleon field in this scenario.}
\label{pl}
\end{figure}

It is straightforward to rewrite the effective mass in terms of the number of e-folds $N$ and evaluate it in terms of the solutions we obtained in the previous sections. Figure \ref{ci} and \ref{pl} show that the effective mass is much bigger than the Hubble scale during inflation. More than that, in both cases the chameleon coupling effectively makes the entropy modes more and more massive during inflation, strongly suppressing its effects. 

One can ask what happens with the sub-Hubble modes. Assuming an usual Bunch-Davies vacuum initial condition, $\zeta$ and $\delta s$ initially have amplitudes of the same order. Using the time scales for the variation of the perturbations, $\Dot{\zeta} \sim H \zeta$ and $\delta\Dot{ s} \sim m_{\text{eff}}\delta s$, one can see from (\ref{entropy}) that the feeding of entropy modes by adiabatic perturbations is negligible  also in small scales. Therefore, in a semi-classical point of view (where $\zeta$ and $\delta s$ are classical fields with quantum initial conditions), the contributions for the power spectrum of adiabatic perturbations are of the order $\mathcal{O}(H^2/m^2_{\text{eff}})$ and can be safely neglected in our model.

From the above discussion, we only need to solve the homogeneous equation for the adiabatic modes 
\begin{align}
    \Ddot{\zeta}+ \left(3H - 2\frac{\Dot{H}}{H}+ \frac{\Ddot{H}}{\Dot{H}}\right)\Dot{\zeta} + \frac{k^2}{a^2}\zeta = 0,
\end{align}
which leads to the usual relation for the power spectra of the perturbations
\begin{align}
    n_s-1 = 2\eta -4\varepsilon,
\end{align}
with the $\varepsilon$ and $\eta$ written in terms of $H$ as functions of $N$:
\begin{align}
    \varepsilon(N) &= -\frac{H'(N)}{H(N)}\nonumber\\
    \eta(N) &= \varepsilon - \frac{1}{2\varepsilon}\varepsilon'(N)\nonumber\\
    &= -\frac{1}{2}\left(\frac{H''(N)}{H'(N)}+ \frac{H'(N)}{H(N)} \right).
\end{align}
Also, since tensor perturbations are independent of scalar ones, we can use:
\begin{align}
    r = 16\varepsilon(N).
\end{align}

As we have previously discussed, in order to leave the attractor phase we need to choose a time to end inflation. We will use the time inflation would end if we didn't have a chameleon coupling for both models. This will also be important to compare how the coupling with chameleon change the observables values for the conformal inflation model.

In the conformal inflation potential case, the tilt and scalar to tensor ration is then given by
\begin{align}
    n_s - 1 = 0.9743, \quad r = 0.0508,
\end{align}
for $50$ e-folds (everything evaluated at $N=201$) and
\begin{align}
    n_s - 1 = 0.9765, \quad r = 0.0421,
\end{align}
for $60$ e-folds (everything evaluated at $N=191$).

For the inverse power law potential case, we have
\begin{align}
    n_s - 1 = 0.9712, \quad r = 0.0292,
\end{align}
for $50$ e-folds (everything evaluated at $N=49$) and
\begin{align}
    n_s - 1 = 0.9742, \quad r = 0.0229,
\end{align}
for $60$ e-folds (everything evaluated at $N=39$).

\begin{figure}
   \centering
    \includegraphics[width=0.9\linewidth]{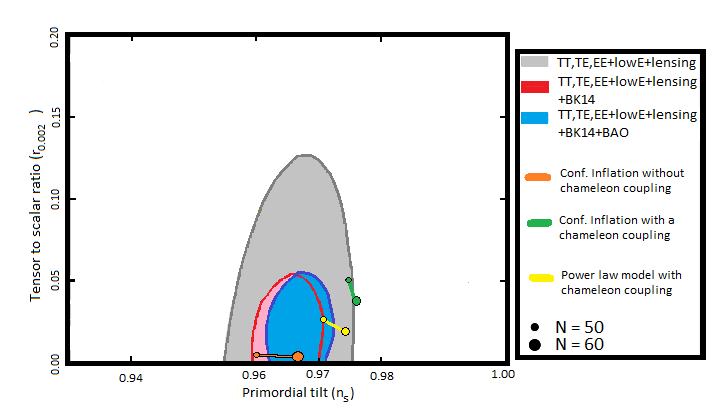}
    \caption{The chameleon coupling to the conformal inflation model almost take its observables outside the allowed region by Planck satelite measurents \cite{Akrami:2018odb}. For the power law model it makes inflation lasts enough e-folds and the observables lie inside the region allowed by observations.}
    \label{planck}
\end{figure}

It is interesting to compare these results with the pure conformal inflation class of models, where for 50 e-folds we get $n_s=0.9600$, while for 60 e-folds $n_s = 0.9667$ (see Figure \ref{planck}). We see that the coupling of the chameleon with non-relativistic matter shifts the values of cosmological observables from the sweet spot of Planck data to almost out of the region allowed by the observations in the conformal inflation case but they still lie inside the region allowed by observations for the power law model considered here.

\section{Conclusions}

In this paper we have investigated the possibility that the inflaton in conformal inflation models is also a chameleon, containg a coupling
$\rho_mF(\varphi)=\rho_m e^{-c(\varphi-\varphi_0)}$
to the energy density $\rho_m$ of some heavy non-relativistic particles present during inflation.

Depending on the sign of $c$ in the exponent of the coupling, we have found two different behaviours. In the $c<0$ case we could
introduce a long period of almost kinetic domination, with an attractor-like behaviour with $1-\Omega_{\rm kin,\varphi}\ll 1$, that precedes the inflationary
phase, and thus shortens it. In the $c>0$ case we found that there are attractor-like behaviours possible, one of which corresponds to sitting at the
instantaneous minimum $\varphi_{\rm min}(N)$ of the effective potential $V_{\rm eff}(\varphi,N)=V(\varphi)+\rho_X(\varphi,N)$. We have
shown the modifications of the CMBR inflationary observables $n_s$ and $r$ after proving the equivalence between our equations of motion and the ones obtained by the microscopic description presented in \cite{DiMarco:2002eb}. We have checked that the second field is heavier than the Hubble scale and so does not generated entropy modes during inflation. More than that, the chameleon coupling with the second scalar field strong suppress the growth of entropy modes during inflation by increasing its effective mass. For conformal inflation models we have found that the presence of non-relativistic matter coupled to a chameleon shifts the value
of $n_s$ and $r$ from the sweet spot of Planck data to almost out of the region allowed region by the data. For the inverse power law case
we have shown that the coupling with chameleon extends the period of inflation and the values for the observables lie in the region allowed by observations.


{\bf Note added}. After our paper was first posted on the arXiv, we became aware of the paper \cite{Cespedes:2015jga}, that also deals with inflation 
coupled to norelativistic matter, though mostly from supergravity embeddings. Their conclusions seem somewhat different, probably because of the 
differences in models, and their specific initial conditions, but they also find that inflation is modified by the chameleon coupling.

\section*{Acknowledgements}

We thank professor Robert Brandenberger for useful discussions. The work of HN is supported in part by CNPq grant 304006/2016-5 and FAPESP grant 2014/18634-9. HN would also
like to thank the ICTP-SAIFR for their support through FAPESP grant 2016/01343-7. RC thanks ICTP-SAIFR for partial support and additional
hospitality and  full financial support by the SARChI NRF grantholder. HB is supported by CAPES.
AW gratefully acknowledges financial support from the South African Research Chairs Initiative of the NRF and the DST. HB and RC are would like to thank McGill University for hospitality during part of this project. Any opinion, finding and conclusion or recommendation expressed in this material is that of the authors and the NRF does not accept any liability in this regard.

\bibliography{confinfcham}
\bibliographystyle{utphys}

\end{document}